\documentclass[aps,superscriptaddress,eqsecnum,nofootinbib,preprintnumbers]{revtex4}
\usepackage{graphicx,epsfig}
\usepackage{amsmath}
\usepackage {amssymb}
\usepackage{subfigure}

\usepackage{relsize}

\newcommand{\be}{\begin{equation}}
\newcommand{\ee}{\end{equation}}
\newcommand{\bea}{\begin{eqnarray}}
\newcommand{\eea}{\end{eqnarray}}
\newcommand{\nn}{\nonumber}

\begin{document}

\title{

The complete Brans-Dicke theories
}

\author{Georgios Kofinas}
\email{gkofinas@aegean.gr} \affiliation{Research Group of Geometry,
Dynamical Systems and Cosmology\\
Department of Information and Communication Systems Engineering\\
University of the Aegean, Karlovassi 83200, Samos, Greece}

\begin{abstract}

Given that the simple wave equation of Brans-Dicke theory for the scalar field is preserved,
we have investigated, through exhaustively analyzing the Bianchi identities, the consistent
theories which violate the exact energy conservation equation. It is found that only three
theories exist which are unambiguously determined from consistency, without imposing
arbitrary functions by hand. Each of these theories possesses a specific interaction term
which controls the energy exchange between the scalar field and ordinary matter. The theories
contain new parameters (integration constants from the integration procedure) and when these
are switched-off, Brans-Dicke theory emerges. As usually, the vacuum theories can be defined
from the complete Brans-Dicke theories when the matter energy-momentum tensor vanishes.

\end{abstract}

\maketitle

\section{Introduction}
\label{intro}

Brans-Dicke gravity theory introduces an additional long-range scalar field $\phi$ besides the
metric tensor $g_{\mu\nu}$ of spacetime and is considered a viable alternative to General Relativity,
one which respects Mach's principle. The effective gravitational constant $G$ is proportional to the
inverse of the scalar field. The scalar field does not exert any direct influence
on matter, its only role is that of participating in the field equations that determine the geometry
of the spacetime. In the following, we will present three uniquely defined generalizations of
Brans-Dicke theory withdrawing this constraint. The word ``uniquely'' means that the theories to be
presented, not only are they consistent generalizations of Brans-Dicke, but they are also natural,
arising without arbitrarily selecting free functions, therefore, they are the only theoretically
favored and interesting theories.

\section{The complete theories with $\Box\phi=4\pi\lambda\mathcal{T}$}
\label{complete}

The cosmic relation $H_{0}^{2}\sim G_{\!N}\rho_{0}$, where $H_{0},\rho_{0}$ are the present Hubble and
energy density parameters and $G_{\!N}$ the Newton's constant, is also written as
$\frac{G_{\!N}^{-1}}{R_{H,0}^{2}}\sim\rho_{0}$, where $R_{H,0}\sim H_{0}^{-1}$ is the present cosmic horizon
size. An equation of the form $\Box\phi\sim\rho$, where $\phi\sim G^{-1}$, is the simplest
spacetime dependent realization of the previous relation and was adopted in Brans-Dicke theory
in order to incorporate Mach's principle. The dimensions of $\phi$ are mass squared, while
a scalar in a standard kinetic term has just a mass dimension.
We consider four spacetime dimensions with metric $g_{\mu\nu}$, a scalar field
$\phi$, while the rest energy-momentum tensor is denoted by $\mathcal{T}^{\mu}_{\,\,\,\nu}$
(e.g. the standard matter-radiation in the case of cosmology) and
$\mathcal{T}=\mathcal{T}^{\mu}_{\,\,\,\,\mu}$.
Equation $\Box\phi=4\pi\lambda\mathcal{T}$ will be adopted in our construction, where
$\lambda\neq 0$ is the standard Brans-Dicke dimensionless parameter. We present in the following
three gravity theories at the level of the equations of motion.
\newline
\newline
\underline{\textit{First theory:}}
\begin{eqnarray}
&&\!\!\!\!\!\!\!G^{\mu}_{\,\,\,\nu}\!=\!\frac{8\pi}{\phi}
(T^{\mu}_{\,\,\,\nu}+\mathcal{T}^{\mu}_{\,\,\,\,\nu})
\label{elp}\\
&&\!\!\!\!\!\!\!T^{\mu}_{\,\,\,\nu}\!=\!\frac{\phi}{2\lambda(\nu\!+\!8\pi\phi^{2})^{2}}\Big{\{}
2\big[(1\!+\!\lambda)\nu\!+\!4\pi(2\!-\!3\lambda)\phi^{2}\big]\phi^{;\mu}\phi_{;\nu}
\!-\!\big[(1\!+\!2\lambda)\nu\!+\!4\pi(2\!-\!3\lambda)\phi^{2}\big]\delta^{\mu}_{\,\,\,\nu}
\phi^{;\rho}\phi_{;\rho} \Big{\}}
\!+\!\frac{\phi^{2}}{\nu\!+\!8\pi\phi^{2}}
\big(\phi^{;\mu}_{\,\,\,\,;\nu}\!-\!\delta^{\mu}_{\,\,\,\nu}\Box\phi\big)
\nn\\
\label{utd}\\
&&\!\!\!\!\!\!\!\Box\phi\!=\!4\pi\lambda\mathcal{T}\label{lrs}\\
&&\!\!\!\!\!\!\!\mathcal{T}^{\mu}_{\,\,\,\,\nu;\mu}\!=\!\frac{\nu}{\phi(\nu\!+\!8\pi\phi^{2})}
\mathcal{T}^{\mu}_{\,\,\,\,\nu}\phi_{;\mu}\,,
\label{idj}
\end{eqnarray}
where $\nu$ is an arbitrary parameter with dimensions mass to the fourth and arises as an
integration constant from the integration procedure.
\newline
\newline
\underline{\textit{Second theory:}}
\begin{eqnarray}
G^{\mu}_{\,\,\,\nu}\!\!&=&\!\!\frac{8\pi}{\phi}
(T^{\mu}_{\,\,\,\nu}+\mathcal{T}^{\mu}_{\,\,\,\,\nu})
\label{elpz}\\
T^{\mu}_{\,\,\,\nu}\!\!&=&\!\!\frac{2\!-\!3\lambda\!-\!4\mu}{16\pi\lambda\phi}
\Big(\phi^{;\mu}\phi_{;\nu}\!-\!\frac{1}{2}\delta^{\mu}_{\,\,\,\nu}
\phi^{;\rho}\phi_{;\rho} \Big)
\!+\!\frac{1}{8\pi}\big(\phi^{;\mu}_{\,\,\,\,;\nu}\!-\!\delta^{\mu}_{\,\,\,\nu}\Box\phi\big)
\label{utdz}\\
\Box\phi\!\!&=&\!\!4\pi\lambda\mathcal{T}\label{lrsz}\\
\mathcal{T}^{\mu}_{\,\,\,\,\nu;\mu}\!\!&=&\!\!\frac{\mu}{\phi}\mathcal{T}\phi_{;\nu}\,,
\label{idjz}
\end{eqnarray}
where $\mu$ is an arbitrary dimensionless parameter and arises as an
integration constant from the integration procedure.
\newline
\newline
\underline{\textit{Third theory:}}
\begin{eqnarray}
G^{\mu}_{\,\,\,\nu}\!\!&=&\!\!\frac{8\pi}{\phi}
(T^{\mu}_{\,\,\,\nu}+\mathcal{T}^{\mu}_{\,\,\,\,\nu})
\label{kst}\\
T^{\mu}_{\,\,\,\nu}\!\!&=&\!\!\frac{2\!-\!3\lambda\!-\!8\sigma}{16\pi\lambda\phi}
\Big(\phi^{;\mu}\phi_{;\nu}\!-\!\frac{1}{2}\delta^{\mu}_{\,\,\,\nu}
\phi^{;\rho}\phi_{;\rho} \Big)
\!+\!\frac{1}{8\pi}\phi^{;\mu}_{\,\,\,\,;\nu}-\frac{\lambda\!+\!2\sigma\!+\!2\eta\phi^{2}}
{8\pi\lambda}\delta^{\mu}_{\,\,\,\nu}\Box\phi
\label{wth}\\
\Box\phi\!\!&=&\!\!4\pi\lambda\mathcal{T}\label{jdh}\\
\mathcal{T}^{\mu}_{\,\,\,\,\nu;\mu}\!\!&=&\!\!(\sigma\!+\!\eta\phi^{2})\mathcal{T}_{;\nu}\,,
\label{iwe}
\end{eqnarray}
where $\sigma$ is an arbitrary dimensionless parameter, $\eta$ is an arbitrary parameter with
dimensions mass to the minus fourth, and both these parameters arise as integration constants from
the integration procedure. Note that this theory, contrary to the other two, does not have
derivative of the scalar field on the right hand hand of the conservation equation (\ref{iwe}),
so even for a slowly varying $\phi$, the geodesic equation is not recovered.

All the above three systems have the following properties:
\newline
(i) The right hand side of equation (\ref{elp}), (\ref{elpz}), (\ref{kst}) is consistent with the
Bianchi identities, i.e. it is covariantly conserved on-shell, and therefore the systems
(\ref{elp})-(\ref{idj}), (\ref{elpz})-(\ref{idjz}), (\ref{kst})-(\ref{iwe}) are well-defined.
\newline
(ii) The first system for $\nu=0$, the second system for $\mu=0$, and the third system for
$\sigma=\eta=0$ all reduce to the standard Brans-Dicke theory \cite{Brans:1961sx}
(in units with $c=1$)
\begin{eqnarray}
G^{\mu}_{\,\,\,\nu}\!\!&=&\!\!\frac{8\pi}{\phi}(T^{\mu}_{\,\,\,\nu}+\mathcal{T}^{\mu}_{\,\,\,\,\nu})
\label{kns}\\
T^{\mu}_{\,\,\,\nu}\!\!&=&\!\!\frac{2\!-\!3\lambda}{16\pi\lambda\phi}\Big(
\phi^{;\mu}\phi_{;\nu}\!-\!\frac{1}{2}\delta^{\mu}_{\,\,\,\nu}
\phi^{;\rho}\phi_{;\rho} \Big)
\!+\!\frac{1}{8\pi}\big(\phi^{;\mu}_{\,\,\,\,;\nu}\!-\!\delta^{\mu}_{\,\,\,\nu}\Box\phi\big)
\label{qqd}\\
\Box\phi\!\!&=&\!\!4\pi\lambda\mathcal{T}\label{lwn}\\
\,\,\mathcal{T}^{\mu}_{\,\,\,\,\nu;\mu}\!\!&=&\!\!0\,.
\label{jrk}
\end{eqnarray}
(iii) Given that $T^{\mu}_{\,\,\,\nu}$ is constructed from terms each of which involves
two derivatives of one or two $\phi$ fields, and $\phi$ itself, and given the equation
(\ref{lrs}), it is shown that the right hand side of the conservation equation is a linear
combination of the quantities $\mathcal{T}^{\mu}_{\,\,\,\,\nu}\phi_{;\mu}$,
$\mathcal{T}\phi_{;\nu}$, $\mathcal{T}_{;\nu}$ with undetermined functions of $\phi$ as
coefficients. This means that any consistent generalization of Brans-Dicke will contain
one or two arbitrary functions of $\phi$ which have to be selected in an ad-hoc way in
order to close the system, and any such theory
loses naturalness. The systems (\ref{elp})-(\ref{idj}), (\ref{elpz})-(\ref{idjz}),
(\ref{kst})-(\ref{iwe}) are the unique Einstein-like theories which can be constructed
naturally, without selecting arbitrary functions by hand.
In other words, a conservation equation of the form
$\mathcal{T}^{\mu}_{\,\,\,\,\nu;\mu}\sim\mathcal{T}^{\mu}_{\,\,\,\,\nu}\phi_{;\mu}$
describes a specific physical interaction and
necessarily implies the theory (\ref{elp})-(\ref{idj}). Similarly an equation
$\mathcal{T}^{\mu}_{\,\,\,\,\nu;\mu}\sim\mathcal{T}\phi_{;\nu}$ picks up the theory
(\ref{elpz})-(\ref{idjz}), while finally an equation
$\mathcal{T}^{\mu}_{\,\,\,\,\nu;\mu}\sim\mathcal{T}_{;\nu}$ provides the theory
(\ref{kst})-(\ref{iwe}).
Therefore, the above three systems can be called completions of Brans-Dicke theory.

It is obvious that the previous generalizations of Brans-Dicke theory are possible because
$\mathcal{T}^{\mu}_{\,\,\,\,\nu}$ is no longer strictly conserved, but modified conservation laws
hold (for other approaches with modified equations of motion see \cite{Smalley:1974gn},
\cite{Clifton:2006vm}).
The matter couples directly to $\phi$; for example in cosmology interacting fluids are currently
studied extensively \cite{Zimdahl:2012mj}, however, usually the interaction chosen is ad-hoc and
does not arise by any physical theory. Since in the action of Brans-Dicke gravity the scalar
field is non-minimally coupled to the curvature, the same mechanism could also lead to a coupling
between the scalar and matter fields, as happens here. Various studies have analyzed
the exchange of energy from ordered motion by entropy generation due to bulk viscosity, direct decay
or non-adiabatic processes from a scalar field \cite{Barrow:1988yc, Geyer:1999kz}.
For example, in \cite{Clifton:2006vm} a modified wave equation for $\phi$ was considered in
Brans-Dicke theory, which therefore does not fall into our consideration. Of course, the
parameters appeared in the above theories should be such that the equivalence principle is not
violated at the ranges
that it has been tested (equivalence principle becomes fragile also in Brans-Dicke theory against
quantum effects \cite{Cho:1994qv}). Chameleon mechanism \cite{Khoury:2003aq} may contribute to this
direction, since the effective mass of $\phi$ can become density dependent, so a large
effective mass may be acquired in solar scale hiding local experiments, while at cosmological scales
$\phi$ can be effectively light providing cosmological modifications. An alternative for the recovery
of General Relativity in the regions of high energy is the introduction of non-linear
self interactions for $\phi$ through self screening (Vainshtein) mechanisms \cite{Vainshtein:1972sx}.
Since the validity of the
universality of free fall at cosmological scales has not been tested directly, another option is that
the portion of baryonic matter inside $\mathcal{T}^{\mu}_{\,\,\,\,\nu}$ is separately conserved,
while the dark matter of cosmology obeys the non-conservation laws; in this case,
$\mathcal{T}^{\mu}_{\,\,\,\,\nu}$ in $\mathcal{T}^{\mu}_{\,\,\,\,\nu;\mu}$ corresponds to
the dark matter energy-momentum tensor, while all other $\mathcal{T}^{\mu}_{\,\,\,\,\nu}$'s in
the equations are the total ones (baryonic plus dark matter).
Note that the strength of the interaction in (\ref{lrs}) is
controlled by $\lambda$, while in (\ref{idj}), (\ref{idjz}), (\ref{iwe}) by $\nu$, $\mu$ and
$\sigma,\eta$ respectively. Especially equation (\ref{idj}) is also written as
$(\phi^{-1}\!\sqrt{|\nu\!+\!8\pi\phi^{2}|}\,\mathcal{T}^{\mu}_{\,\,\,\,\nu})_{;\mu}=0$ in the form
of the exact conservation of a redefined energy-momentum tensor, while the
conservation equation of $T^{\mu}_{\,\,\,\nu}$ comes from the combination of (\ref{elp}), (\ref{idj})
to be $T^{\mu}_{\,\,\,\nu;\mu}=(T^{\mu}_{\,\,\,\nu}+\frac{8\pi\phi^{2}}{\nu+8\pi\phi^{2}}
\mathcal{T}^{\mu}_{\,\,\,\,\nu})\frac{\phi_{;\mu}}{\phi}$. Similarly from (\ref{elpz}),
(\ref{idjz}) it arises $T^{\mu}_{\,\,\,\nu;\mu}=
(T^{\mu}_{\,\,\,\nu}+\mathcal{T}^{\mu}_{\,\,\,\nu}-\mu\mathcal{T}\delta^{\mu}_{\,\,\,\nu})
\frac{\phi_{;\mu}}{\phi}$, while from (\ref{kst}), (\ref{iwe}) it is
$T^{\mu}_{\,\,\,\nu;\mu}=(T^{\mu}_{\,\,\,\nu}+\mathcal{T}^{\mu}_{\,\,\,\nu})
\frac{\phi_{;\mu}}{\phi}-(\sigma\!+\!\eta\phi^{2})\mathcal{T}_{;\nu}$. Of course, in view of the Bianchi
identities, equation (\ref{idj}) can be derived from (\ref{elp})-(\ref{lrs}) and similarly for
the other systems.

The Brans-Dicke theory can also be formulated in terms of an action solely based on dimensional
arguments, with the matter Lagrangian being minimally coupled. However, if interactions of the matter
Lagrangian with the scalar field are allowed, then plenty of actions could be constructed with limit
the Brans-Dicke action in the absence of interactions. The number of such actions becomes even larger
in the presence of Newton's constant $G_{\!N}$ or a new massive or massless scale like
$\nu,\mu,\sigma,\eta$.
On the opposite, the derivation of the interacting theory through the equations of motion provides a
uniqueness prescription, at least when a single interaction is present.
Of course, an immediate question which arises is what are the corresponding actions, if any,
which give the previous systems.

In the decoupling limit $\lambda\rightarrow 0$ of $\mathcal{T}^{\mu}_{\,\,\,\,\nu}$ and $\phi$,
it is expected that Brans-Dicke theory goes over to the Einstein theory. However, in this limit
equation (\ref{qqd}) formally does not make sense; moreover in this limit, a solution of the system
(\ref{kns})-(\ref{jrk}) does not always reduce to a solution of General
Relativity with the same $\mathcal{T}^{\mu}_{\,\,\,\,\nu}$ \cite{Romero:1992bu}. On the other
hand, the first theory is meaningful in this limit. Indeed,
the full equation (\ref{utd}) makes sense in this limit, since redefining the integration
constant $\nu$, e.g. as $\nu=\lambda^{-2}\nu'$, we get from (\ref{elp})-(\ref{idj}) respectively
$G^{\mu}_{\,\,\,\nu}=8\pi\phi^{-1}\mathcal{T}^{\mu}_{\,\,\,\,\nu}$, $T^{\mu}_{\,\,\,\nu}\equiv 0$,
$\Box\phi=0$, $(\phi^{-1}\mathcal{T}^{\mu}_{\,\,\,\,\nu})_{;\mu}=0$, which is Einstein-like gravity with
a well-defined varying gravitational constant $G\sim\phi^{-1}$ (the constant $\phi$ solution can
be further picked up to get Einstein gravity).
For the second theory it cannot be made $T^{\mu}_{\,\,\,\nu}\equiv 0$ for $\lambda\rightarrow 0$
even redefining the integration constant and the theory does not become Einstein-like.
For the third theory the limit $\lambda\rightarrow 0$ cannot be considered at all, even redefining
the integration constants.

\section{Proof of the statements}
\label{proof}

To prove the previous statements we start with the Bianchi identities
\begin{equation}
G^{\mu}_{\,\,\,\nu}
\phi_{;\mu}-8\pi T^{\mu}_{\,\,\,\nu;\mu}=8\pi\mathcal{T}^{\mu}_{\,\,\,\,\nu;\mu}
\label{eis}
\end{equation}
arising from (\ref{elp}), with $\phi$ playing the role of inverse gravitational parameter.
Equation (\ref{eis}) is also written as
\begin{equation}
T^{\mu}_{\,\,\,\nu;\mu}-\frac{1}{\phi}T^{\mu}_{\,\,\,\nu}\phi_{;\mu}=
\frac{1}{\phi}\mathcal{T}^{\mu}_{\,\,\,\,\nu}\phi_{;\mu}-\mathcal{T}^{\mu}_{\,\,\,\,\nu;\mu}\,.
\label{wjd}
\end{equation}
The most general symmetric tensor that can be built up from terms each of which involves two
derivatives of one or two $\phi$ fields, and $\phi$ itself, is
\begin{equation}
T^{\mu}_{\,\,\,\nu}=A(\phi)\phi^{;\mu}\phi_{;\nu}+B(\phi)\delta^{\mu}_{\,\,\,\nu}\phi^{;\rho}\phi_{;\rho}
+C(\phi)\phi^{;\mu}_{\,\,\,\,;\nu}+E(\phi)\delta^{\mu}_{\,\,\,\nu}\Box{\phi}\,.
\label{jwo}
\end{equation}
Using the Einstein equation (\ref{elp}), the form of $T^{\mu}_{\,\,\,\nu}$ from (\ref{jwo}), and
the identity $R^{\mu}_{\,\,\,\nu}\phi_{;\mu}=\Box{(\phi_{;\nu})}-(\Box\phi)_{;\nu}$\,, we express
$\Box{(\phi_{;\nu})}$ as
\begin{equation}
\Box{(\phi_{;\nu})}=(\Box\phi)_{;\nu}+\frac{4\pi}{\phi}(A\!-\!2B)\phi^{;\mu}\phi_{;\mu}\phi_{;\nu}
-\frac{4\pi}{\phi}(C\!+\!2E)\phi_{;\nu}\Box{\phi}+\frac{8\pi}{\phi}C\phi^{;\mu}_{\,\,\,\,;\nu}
\phi_{;\mu}+\frac{4\pi}{\phi}(2\mathcal{T}^{\mu}_{\,\,\,\,\nu}\!-\!\mathcal{T}\delta^{\mu}_{\,\,\,\nu})
\phi_{;\mu}\,.
\label{lwi}
\end{equation}
The Bianchi identities (\ref{wjd}), differentiating $T^{\mu}_{\,\,\,\nu}$ from (\ref{jwo})
and using (\ref{lwi}), become
\begin{eqnarray}
&&\!\!\!\!\!\!\!\!\!\!\!\!\!\!\!\!\!
\Big[A'\!+\!B'\!+\!\frac{4\pi}{\phi}C(A\!-\!2B)\!-\!\frac{1}{\phi}(A\!+\!B)\Big]
\phi^{;\mu}\phi_{;\mu}\phi_{;\nu}+\Big[A\!+\!E'\!-\!\frac{4\pi}{\phi}C(C\!+\!2E)\!-\!\frac{1}{\phi}E\Big]
\phi_{;\nu}\Box{\phi}
\nn\\
&&\!\!\!\!\!\!\!\!\!\!\!\!\!\!\!\!\!
+\Big[A\!+\!2B\!+\!C'\!+\!\frac{8\pi}{\phi}C^{2}\!-\!\frac{1}{\phi}C\Big]
\phi^{;\mu}_{\,\,\,;\nu}\phi_{;\mu}+(C\!+\!E)(\Box\phi)_{;\nu}
\!+\!\Big(\mathcal{T}^{\mu}_{\,\,\,\,\nu;\mu}-\frac{1\!-\!8\pi C}{\phi}\mathcal{T}^{\mu}_{\,\,\,\,\nu}
\phi_{;\mu}-\frac{4\pi}{\phi}C\mathcal{T}\phi_{;\nu}\Big)=0\,,
\label{ask}
\end{eqnarray}
where a prime denotes differentiation with respect to $\phi$.
Now, replacing $\mathcal{T}$ from (\ref{lrs}) in (\ref{ask}) we get the Bianchi identities in the form
\begin{eqnarray}
&&\!\!\!\!\!\!\!\!\!\!\!\!\!\!\!\!\!
\Big[A'\!+\!B'\!+\!\frac{4\pi}{\phi}C(A\!-\!2B)\!-\!\frac{1}{\phi}(A\!+\!B)\Big]
\phi^{;\mu}\phi_{;\mu}\phi_{;\nu}+\Big[A\!+\!E'\!-\!\frac{4\pi}{\phi}C(C\!+\!2E)\!-\!\frac{1}{\phi}E
-\frac{1}{\lambda \phi}C\Big]\phi_{;\nu}\Box{\phi}
\nn\\
&&\!\!\!\!\!\!\!\!\!\!\!\!\!\!\!\!\!
+\Big[A\!+\!2B\!+\!C'\!+\!\frac{8\pi}{\phi}C^{2}\!-\!\frac{1}{\phi}C\Big]
\phi^{;\mu}_{\,\,\,;\nu}\phi_{;\mu}+(C\!+\!E)(\Box\phi)_{;\nu}
\!+\!\Big(\mathcal{T}^{\mu}_{\,\,\,\,\nu;\mu}-\frac{1\!-\!8\pi C}{\phi}\mathcal{T}^{\mu}_{\,\,\,\,\nu}
\phi_{;\mu}\Big)=0\,.
\label{ope}
\end{eqnarray}
Note that in the zero matter limit, $\mathcal{T}^{\mu}_{\,\,\,\,\nu}=0$, the system should still be
meaningful with the same $A,B,C,E$, and a vacuum theory should be defined. Since in that case it is
$\Box\phi=0$, equation (\ref{ope}) implies the vanishing of the coefficients of
$\phi^{;\mu}\phi_{;\mu}\phi_{;\nu}$ and $\phi^{;\mu}_{\,\,\,;\nu}\phi_{;\mu}$
so that no new equations of motion arise.
Thus, the Bianchi identities (\ref{ope}) become equivalent to the system of equations
\begin{eqnarray}
&&A'\!+\!B'\!+\!\frac{4\pi}{\phi}C(A\!-\!2B)\!-\!\frac{1}{\phi}(A\!+\!B)=0\label{kef1}\\
&&A\!+\!2B\!+\!C'\!+\!\frac{8\pi}{\phi}C^{2}\!-\!\frac{1}{\phi}C=0\label{kej1}\\
&&\mathcal{T}^{\mu}_{\,\,\,\,\nu;\mu}-\frac{1\!-\!8\pi C}{\phi}\mathcal{T}^{\mu}_{\,\,\,\,\nu}
\phi_{;\mu}+\Big[A\!+\!E'\!-\!\frac{4\pi}{\phi}C(C\!+\!2E)\!-\!\frac{1}{\phi}E
-\frac{1}{\lambda \phi}C\Big]\phi_{;\nu}\Box{\phi}+(C\!+\!E)(\Box\phi)_{;\nu}=0\label{dwr1}\,.
\end{eqnarray}
In the conservation equation (\ref{dwr1}), beyond the standard term
$\mathcal{T}^{\mu}_{\,\,\,\,\nu;\mu}$, there are three possible terms of violating the exact
conservation:
$\mathcal{T}^{\mu}_{\,\,\,\,\nu}\phi_{;\mu}$, $\mathcal{T}\phi_{;\nu}$ and $\mathcal{T}_{;\nu}$.
Therefore, the most general possible energy-momentum conservation equation is of the form
$\mathcal{T}^{\mu}_{\,\,\,\,\nu;\mu}=f(\phi)\mathcal{T}^{\mu}_{\,\,\,\,\nu}
\phi_{;\mu}+h(\phi)\mathcal{T}\phi_{;\nu}+m(\phi)\mathcal{T}_{;\nu}$ for appropriate
$f,h,m$. However, equations (\ref{kef1}), (\ref{kej1}) carry the freedom of one arbitrary function
(let's say $C$), and (\ref{dwr1}) contains an extra arbitrary function $E$.
Thus, any $A,B,C,E,\mathcal{T}^{\mu}_{\,\,\,\,\nu}$ satisfying the system (\ref{kef1})-(\ref{dwr1})
is consistent, however, there is an arbitrariness in two undetermined functions of $\phi$.
These two functions should somehow be selected by hand and any such arising theory lacks
naturalness and is deprived from being theoretically favored or interesting.
In the case that only two of the three terms $\mathcal{T}^{\mu}_{\,\,\,\,\nu}\phi_{;\mu}$,
$\mathcal{T}\phi_{;\nu}$, $\mathcal{T}_{;\nu}$ are present in the conservation equation, the
degree of arbitrariness is reduced to one free function. The only way that there is no
arbitrariness is when only one violating term is present and the theory is then naturally
selected. Thus, three unique theories arise by vanishing the two of the three coefficients
of $\mathcal{T}^{\mu}_{\,\,\,\,\nu}\phi_{;\mu}$, $\mathcal{T}\phi_{;\nu}$, $\mathcal{T}_{;\nu}$
in (\ref{dwr1}). The first such theory will have a conservation equation of the form
$\mathcal{T}^{\mu}_{\,\,\,\,\nu;\mu}=f(\phi)\mathcal{T}^{\mu}_{\,\,\,\,\nu}\phi_{;\mu}$ for suitable
$f(\phi)$, along with the appropriate $T^{\mu}_{\,\,\,\,\nu}$. Similarly, the second one will have
$\mathcal{T}^{\mu}_{\,\,\,\,\nu;\mu}=h(\phi)\mathcal{T}\phi_{;\nu}$
and the third $\mathcal{T}^{\mu}_{\,\,\,\,\nu;\mu}=m(\phi)\mathcal{T}_{;\nu}$. We are going to
find below these general three theories.
Note that in order for a theory to generalize Brans-Dicke, some extra parameters/integration
constants should be present, which for appropriate values reduce the system to the Brans-Dicke
one.

{\textit{First theory :
$\mathcal{T}^{\mu}_{\,\,\,\,\nu;\mu}\sim\mathcal{T}^{\mu}_{\,\,\,\,\nu}\phi_{;\mu}$.}}
This case corresponds to vanishing both coefficients of
$\phi_{;\nu}\Box{\phi}$ and $(\Box\phi)_{;\nu}$ in (\ref{dwr1}). Therefore, we have the
following system of differential equations
\begin{eqnarray}
&&A'\!+\!B'\!+\!\frac{4\pi}{\phi}C(A\!-\!2B)\!-\!\frac{1}{\phi}(A\!+\!B)=0\label{kef}\\
&&A\!+\!2B\!+\!C'\!+\!\frac{8\pi}{\phi}C^{2}\!-\!\frac{1}{\phi}C=0\label{kej}\\
&&A\!+\!E'\!-\!\frac{4\pi}{\phi}C(C\!+\!2E)\!-\!\frac{1}{\phi}E
-\frac{1}{\lambda \phi}C=0\label{dhk}\\
&&C\!+\!E=0\label{jdk}\\
&&\mathcal{T}^{\mu}_{\,\,\,\,\nu;\mu}=\frac{1\!-\!8\pi C}{\phi}\mathcal{T}^{\mu}_{\,\,\,\,\nu}
\phi_{;\mu}\label{kev}\,.
\end{eqnarray}
Due to these equations, the Bianchi identities are identically satisfied and a unique
theory will arise without introducing arbitrary functions.
The general solution of the system (\ref{kef})-(\ref{jdk}) will be found next. Using (\ref{jdk})
and replacing (\ref{kej}) with the sum of (\ref{kej}) and (\ref{dhk}), we obtain the equivalent
system
\begin{eqnarray}
&&(A\!+\!B)'\!+\!\frac{4\pi}{\phi}C(A\!-\!2B)\!-\!\frac{1}{\phi}(A\!+\!B)=0\label{jwe}\\
&&A\!+\!B\!+\!\frac{6\pi}{\phi}C^{2}\!-\!\frac{1}{2\lambda\phi}C=0\label{jke}\\
&&C'\!-\!\frac{4\pi}{\phi}C^{2}\!-\!\Big(1\!-\!\frac{1}{\lambda}\Big)\frac{1}{\phi}C\!-\!A=0\label{keb}\\
&&E=-C\label{bsa}\,.
\end{eqnarray}
Defining the quantities $X=A+B$, $Y=A-B$, the system (\ref{jwe})-(\ref{bsa}) gets the form
\begin{eqnarray}
&&X'\!+\!\frac{2\pi}{\phi}C(3Y\!-\!X)\!-\!\frac{1}{\phi}X=0\label{nbd}\\
&&X=-\frac{6\pi}{\phi}C^{2}\!+\!\frac{1}{2\lambda\phi}C\label{maf}\\
&&Y=2C'-\frac{8\pi}{\phi}C^{2}-2\Big(1\!-\!\frac{1}{\lambda}\Big)\frac{1}{\phi}C\!-\!X\label{lah}\\
&&E=-C\label{fsa}\,.
\end{eqnarray}
Differentiating (\ref{maf}) and replacing this derivative, as well as $X$ itself and $Y$ from
(\ref{lah}), into (\ref{nbd}), we get instead of (\ref{nbd}) the following simple differential
equation for $C$
\begin{equation}
C'\!+\!\frac{16\pi}{\phi}C^{2}\!-\!\frac{2}{\phi}C=0\,.
\label{iwj}
\end{equation}
The general solution of equation (\ref{iwj}) is
\begin{equation}
C=\frac{\phi^{2}}{\nu\!+\!8\pi\phi^{2}}\,,
\label{wdf}
\end{equation}
where $\nu$ is integration constant (appeared in (\ref{utd}), (\ref{idj}) as parameter). Substituting
this $C$ into (\ref{fsa}), (\ref{maf}), (\ref{lah}), we obtain $E,X,Y$ as functions of $\phi$, and then
$A(\phi)$, $B(\phi)$ also arise. The result is the energy-momentum tensor appeared in (\ref{utd}).
Finally, plugging $C$ into (\ref{kev}) we obtain equation (\ref{idj}).

{\textit{Second theory : $\mathcal{T}^{\mu}_{\,\,\,\,\nu;\mu}\sim\mathcal{T}\phi_{;\nu}$.}}
This case corresponds to setting
\begin{equation}
C=\frac{1}{8\pi}\,\,\,\,\,\,,\,\,\,\,\,\,C+E=0\label{jwr}
\end{equation}
in (\ref{dwr1}), and therefore equation (\ref{dwr1}) becomes
\begin{equation}
\mathcal{T}^{\mu}_{\,\,\,\,\nu;\mu}+\Big(A+\frac{3}{16\pi\phi}-\frac{1}{8\pi\lambda\phi}\Big)
\phi_{;\nu}\Box{\phi}=0\,.\label{jeq}
\end{equation}
Solving the algebraic equation (\ref{kej1}) for $A$ we find $A=-2B$ and substituting in
(\ref{kef1}) we get
\begin{equation}
B'+\frac{1}{\phi}B=0\,.
\label{qfk}
\end{equation}
The solution of (\ref{qfk}) is
\begin{equation}
B=\frac{\mu_{1}}{\phi}\,,
\label{evo}
\end{equation}
where $\mu_{1}$ is integration constant. The conservation equation (\ref{jeq}) takes the form
(\ref{idjz}), where $\mu=4\pi\lambda\big(2\mu_{1}+\frac{2-3\lambda}{16\pi\lambda}\big)$ is a
redefined integration constant (appeared in (\ref{utdz}), (\ref{idjz}) as parameter).
The energy-momentum tensor (\ref{jwo}) of the scalar field takes the form (\ref{utdz}).
For $\mu=\frac{1}{2}$ this theory reduces to the theory found in \cite{Barber:2002zz} under
the name of self creation cosmology.

{\textit{Third theory : $\mathcal{T}^{\mu}_{\,\,\,\,\nu;\mu}\sim\mathcal{T}_{;\nu}$.}}
This case corresponds to setting
\begin{eqnarray}
&&C=\frac{1}{8\pi}\label{sne}\\
&&A\!+\!E'\!-\!\frac{4\pi}{\phi}C(C\!+\!2E)\!-\!\frac{1}{\phi}E
-\frac{1}{\lambda\phi}C=0\label{rdg}
\end{eqnarray}
in (\ref{dwr1}), and therefore equation (\ref{dwr1}) becomes
\begin{equation}
\mathcal{T}^{\mu}_{\,\,\,\,\nu;\mu}+(C\!+\!E)(\Box\phi)_{;\nu}=0\,.\label{keb}
\end{equation}
Solving the algebraic equation (\ref{kej1}) for $A$ we find $A=-2B$ and substituting in
(\ref{kef1}) we get
\begin{equation}
B'+\frac{1}{\phi}B=0\,.
\label{qfv}
\end{equation}
The solution of (\ref{qfv}) is
\begin{equation}
B=\frac{\sigma_{1}}{\phi}\,,
\label{evo}
\end{equation}
where $\sigma_{1}$ is integration constant. Equation (\ref{rdg}) becomes
\begin{equation}
E'-\frac{2}{\phi}E-\Big(2\sigma_{1}\!+\!\frac{\lambda\!+\!2}{16\pi\lambda}\Big)\frac{1}{\phi}=0,
\label{eby}
\end{equation}
with general solution
\begin{equation}
E=-\frac{2\eta\phi^{2}\!+\!\lambda\!+\!2\sigma}{8\pi\lambda}\,,\label{wto}
\end{equation}
where $\sigma=4\pi\lambda\big(\sigma_{1}+\frac{2-3\lambda}{32\pi\lambda}\big)$ is a redefined
integration constant and $\eta$ is a another integration constant (appeared in (\ref{wth}),
(\ref{iwe}) as parameters).
The conservation equation (\ref{keb}) takes the form (\ref{iwe}),
while the energy-momentum tensor (\ref{jwo}) of the scalar field gives equation (\ref{wth}).

To compare with the standard derivation of Brans-Dicke gravity \cite{Weinberg}, there, the quantity
$R^{\mu}_{\,\,\,\nu}\phi_{;\mu}$ of the Bianchi identities (\ref{eis}) is replaced by
$\Box(\phi_{;\nu})-(\Box\phi)_{;\nu}$, and after use of (\ref{jwo}), (\ref{elp}), equation
(\ref{eis}) becomes
\begin{eqnarray}
&&\Big[A'\!+\!B'\!-\!\frac{1}{2\phi}(A\!+\!4B)\Big]\phi^{;\mu}\phi_{;\mu}\phi_{;\nu}
\!+\!\Big[A\!+\!E'\!-\!\frac{1}{2\phi}(C\!+\!4E)\Big]\phi_{;\nu}\Box{\phi}
\!+\!(A\!+\!2B\!+\!C')\phi^{;\mu}_{\,\,\,\,;\nu}\phi_{;\mu}\nn\\
&&\!+\Big(E+\frac{1}{8\pi}\Big)(\Box{\phi})_{;\nu}\!+\!\Big(C\!-\!\frac{1}{8\pi}\Big)
\Box{(\phi_{;\nu})}\!+\!\Big(\mathcal{T}^{\mu}_{\,\,\,\,\nu;\mu}\!-\!\frac{1}{2\phi}\mathcal{T}
\phi_{;\nu}\Big)=0\,.
\label{owf}
\end{eqnarray}
Replacing $\mathcal{T}$ from (\ref{lrs}), equation (\ref{owf}) becomes
\begin{eqnarray}
&&\Big[A'\!+\!B'\!-\!\frac{1}{2\phi}(A\!+\!4B)\Big]\phi^{;\mu}\phi_{;\mu}\phi_{;\nu}
\!+\!\Big[A\!+\!E'\!-\!\frac{1}{2\phi}(C\!+\!4E)\!-\!\frac{1}{8\pi\lambda\phi}\Big]\phi_{;\nu}\Box{\phi}
\!+\!(A\!+\!2B\!+\!C')\phi^{;\mu}_{\,\,\,\,;\nu}\phi_{;\mu}\nn\\
&&\!+\Big(E+\frac{1}{8\pi}\Big)(\Box{\phi})_{;\nu}\!+\!\Big(C\!-\!\frac{1}{8\pi}\Big)
\Box{(\phi_{;\nu})}\!+\!\mathcal{T}^{\mu}_{\,\,\,\,\nu;\mu}=0\,.
\label{oww}
\end{eqnarray}
The main difference between equations (\ref{ope}) and (\ref{oww}) is that in (\ref{ope}) the
quantity $\Box(\phi_{;\nu})$ has been expressed in terms of the quantity $(\Box\phi)_{;\nu}$,
so these quantities are not independent, while in (\ref{oww}) both quantities are present and
their coefficients vanish separately. This difference in the two approaches occurs not only here in
the matter case, but also in section \ref{vacuum} of the vacuum case.
Vanishing the various coefficients in (\ref{oww}) in order to satisfy the Bianchi identities,
a system of (more) equations arise which is basically algebraic (the sole differential equation is
identically satisfied) with solution the special theory (\ref{kns})-(\ref{jrk}) with no integration
constant. On the contrary, the system of equations (\ref{kef})-(\ref{kev}) is weaker (fewer equations)
and purely differential, so its solution becomes more enlarged (similarly for the other two systems).

Note that a different than (\ref{lrs}) equation of motion for the field $\phi$ would give
different theories. For example, for $\Box\phi+F(\phi)\phi^{;\mu}\phi_{;\mu}=4\pi\lambda\mathcal{T}$,
the coefficient of $\phi^{;\mu}\phi_{;\mu}\phi_{;\nu}$ in (\ref{ope}) would be different.
The interesting case with $\Box\phi=V'(\phi)+4\pi\lambda\mathcal{T}$ would leave equations
(\ref{elp}), (\ref{utd}) unchanged, while on the right hand side of (\ref{idj}) the term
$-\frac{\phi}{\nu+8\pi\phi^{2}}\frac{V'}{\lambda}\phi_{;\nu}$ should be added; in the absence of
matter, $V$ should be just a cosmological constant (alternatively, a self-interacting potential could be
introduced through an action with different results).

\section{Generalized vacuum Brans-Dicke theories}

\label{vacuum}
It is a standard issue in Brans-Dicke gravity (\ref{kns})-(\ref{jrk}) that the vacuum theory
arises by vanishing the matter energy-momentum tensor $\mathcal{T}^{\mu}_{\,\,\,\,\nu}=0$
(for vacuum solutions see \cite{Brans:1961sx, Mahanta:1971pk, Anchordoqui:1997yb}). This means that
the scalar field equation of motion is the free equation $\Box\phi=0$, together with the Einstein
equation (\ref{kns}), (\ref{qqd}). In this system there still appears the parameter $\lambda$
which now does not control any coupling, as was happening in equation (\ref{lwn}) of the matter theory.
However, there is a continuity between the purely vacuum theory (setting
$\mathcal{T}^{\mu}_{\,\,\,\,\nu}=0$ from the beginning) and the vacuum theory which arises as the
above limiting process of the matter theory. Indeed, the consistency of equation (\ref{owf}) with
$\mathcal{T}^{\mu}_{\,\,\,\,\nu}=0$, $\Box\phi=0$, gives a system of three vanishing coefficients
whose integration provides $T^{\mu}_{\,\,\,\nu}$ of (\ref{qqd}) with the difference that now $\lambda$
is an integration constant.

The three complete Brans-Dicke theories (\ref{elp})-(\ref{idj}), (\ref{elpz})-(\ref{idjz}),
(\ref{kst})-(\ref{iwe}) of the previous section in the presence of
extra matter, already define for $\mathcal{T}^{\mu}_{\,\,\,\,\nu}=0$ generalized vacuum
Brans-Dicke theories (setting $\nu=0$ in the first theory, it reduces to the vacuum Brans-Dicke
theory, and similarly setting $\mu=0$ in the second and $\sigma=\eta=0$ in the third theory).
These vacuum theories arising from the limiting process of the matter theories are certainly
the most interesting generalizations of vacuum Brans-Dicke. However, as with the matter theories,
also with the vacuum theories there is an infinity of possibilities respecting the wave equation
$\Box\phi=0$. This cannot be seen, as we explained, from equation
(\ref{owf}). However, it can be seen from equation (\ref{ask}) (before $\mathcal{T}$ is replaced
in terms of $\Box\phi$ as in (\ref{ope})), from where the system of equations (\ref{kef1}),
(\ref{kej1}) arises. The solution of these equations is
\begin{eqnarray}
&&A=2e^{\int(1-16\pi C)\frac{d\phi}{\phi}}\Big[\alpha\!-\!12\pi \!\int \!d\phi\, \frac{C}{\phi}
\Big(C'\!+\!\frac{8\pi}{\phi} C^{2}\!-\!\frac{1}{\phi}C\Big)e^{\int(16\pi C-1)\frac{d\phi}{\phi}}\Big]
+C'\!+\!\frac{8\pi}{\phi} C^{2}\!-\!\frac{1}{\phi}C \label{kefe}\\
&&B=-e^{\int(1-16\pi C)\frac{d\phi}{\phi}}\Big[\alpha\!-\!12\pi \!\int \!d\phi\, \frac{C}{\phi}
\Big(C'\!+\!\frac{8\pi}{\phi} C^{2}\!-\!\frac{1}{\phi}C\Big)e^{\int(16\pi C-1)\frac{d\phi}{\phi}}\Big]
-\Big(C'\!+\!\frac{8\pi}{\phi} C^{2}\!-\!\frac{1}{\phi}C\Big) \label{wefe}
\end{eqnarray}
(with $\alpha$ integration constant) and is parametrized by one free function $C(\phi)$. Solutions
(\ref{kefe}), (\ref{wefe}) can also be used in (\ref{dwr1}), (\ref{jwo}) to describe the infinity
of the matter theories (along with the free $E(\phi)$). Of course, the stress tensors
$T^{\mu}_{\,\,\,\nu}$ of the three matter theories satisfy (\ref{kefe}), (\ref{wefe}).

It is interesting to be able to construct under some prescription consistent non-trivial systems for
$g_{\mu\nu},\phi$ as above with simple equations of motion for $\phi$, for example $\Box\phi=0$, or
some modification $\Box\phi=V'(\phi)$ with $V=\frac{1}{2}m^{2}\phi^{2}$ or $V$ a potential analogous
to the one of the Higgs field of particle physics. Starting on the contrary with an arbitrary
energy-momentum tensor $T^{\mu}_{\,\,\,\nu}$ of $\phi$ on the right hand side of $G^{\mu}_{\,\,\,\nu}$,
will give through the Bianchi identities the equation of motion for $\phi$ which will in general be
complicated. Another option is to assume actions, but still there, the equations of motion for the
scalar field are in general complicated. Of course, there is the problem with the undetermined
function $C$, which is however resolved through the vacuum limit of the matter theories.

In the context of scalar-tensor gravity, it has been constructed \cite{Horndeski:1974wa} the most general
theory with second-order equations of motion which arise from action (see also \cite{Deffayet:2009mn}).
The role of the action is crucial in that it provides an unexpected relation \cite{Hessis} between the
scalar field equation of motion and the divergence of the gravitational equations, thus the scalar
equation is a consequence of the gravity equations. Here, the field equations do not arise
demanding an action and the scalar equation of motion does not obey necessarily such a relation.
Therefore, it is to be seen if a Lagrangian can be constructed to derive the theories
described by (\ref{kefe}), (\ref{wefe}).
In this case, these theories will provide particular sectors of the full Horndeski family.

One way to satisfy the Bianchi identities $(\phi^{-1}T^{\mu}_{\,\,\,\nu})_{;\mu}$ with
$\mathcal{T}^{\mu}_{\,\,\,\nu}=0$ is to satisfy (\ref{ask}) identically requiring stronger
conditions, i.e. demanding the vanishing of the four relevant coefficients. This is quite
different than vanishing the four coefficients in (\ref{ope}). The arising system of equations
consists again of (\ref{kef})-(\ref{jdk}) with the difference that the $\lambda$ term in
(\ref{dhk}) is now absent. This absence leads to the redundancy of one equation.
The vanishing of all coefficients in (\ref{ask}) has the physical
meaning that the arising scalar-tensor theory $G^{\mu}_{\,\,\,\nu}=8\pi\phi^{-1}T^{\mu}_{\,\,\,\nu}$
can be supplemented by any field equation for $\phi$ (containing $\phi,g_{\mu\nu}$)
and not only by $\Box\phi=0$. Unfortunately, there is still here the indeterminacy of one
free function of a different form than in (\ref{kefe}), (\ref{wefe}). Indeed, now
the following energy-momentum comes up
\begin{equation}
T^{\mu}_{\,\,\,\nu}=\Big(C'\!-\!\frac{4\pi}{\phi}C^{2}\!-\!\frac{1}{\phi}C\Big)
\phi^{;\mu}\phi_{;\nu}
\!-\!\Big(C'\!+\!\frac{2\pi}{\phi}C^{2}\!-\!\frac{1}{\phi}C\Big)\delta^{\mu}_{\,\,\,\nu}
\phi^{;\rho}\phi_{;\rho}
\!+\!C\big(\phi^{;\mu}_{\,\,\,\,;\nu}\!-\!\delta^{\mu}_{\,\,\,\nu}\Box\phi\big)\,.
\label{utc}
\end{equation}
The stress tensors $T^{\mu}_{\,\,\,\nu}$ of the three matter theories do not in general belong
in this class. To see the limits of (\ref{utc}), setting
$C=\frac{1}{8\pi}$, we get the -questionable due to (\ref{lwn})- $\lambda\rightarrow\infty$
(or $\omega=\frac{2-3\lambda}{2\lambda}=-\frac{3}{2}$) limit of Brans-Dicke equation (\ref{qqd}).
This limit also arises from the less efficient equation (\ref{owf}) vanishing the various coefficients
and it can be accompanied by any equation of motion for $\phi$.
To give an example of (\ref{utc}), choosing $C=\frac{\phi}{\beta+8\pi\phi}$,
where $\beta$ is an arbitrary parameter, we obtain
\begin{eqnarray}
T^{\mu}_{\,\,\,\nu}\!\!&=&\!\!-\frac{12\pi\phi}{(\beta\!+\!8\pi\phi)^{2}}
\Big(\phi^{;\mu}\phi_{;\nu}\!-\!\frac{1}{2}\delta^{\mu}_{\,\,\,\nu}
\phi^{;\rho}\phi_{;\rho}\Big)\!+\!\frac{\phi}{\beta\!+\!8\pi\phi}
\big(\phi^{;\mu}_{\,\,\,\,;\nu}\!-\!\delta^{\mu}_{\,\,\,\nu}\Box\phi\big)\label{kdr}\\
\!\!&=&\!\!-\frac{3}{16\pi\varphi}\Big(1\!-\!\frac{\beta}{8\pi\varphi}\Big)
\Big(\varphi^{;\mu}\varphi_{;\nu}\!-\!\frac{1}{2}\delta^{\mu}_{\,\,\,\nu}
\varphi^{;\rho}\varphi_{;\rho}\Big)\!+\!\frac{1}{8\pi}\Big(1\!-\!\frac{\beta}{8\pi\varphi}\Big)
\big(\varphi^{;\mu}_{\,\,\,\,;\nu}\!-\!\delta^{\mu}_{\,\,\,\nu}\Box\varphi\big)\,,
\label{kel}
\end{eqnarray}
where $\varphi=\phi+\frac{\beta}{8\pi}$. Actually this $C$ is the only
choice in order to form the combination
$\phi^{;\mu}\phi_{;\nu}\!-\!\frac{1}{2}\delta^{\mu}_{\,\,\,\nu}
\phi^{;\rho}\phi_{;\rho}$, and therefore $\beta$ appears as an integration constant.
Setting $\beta=0$ in (\ref{kdr}), once again we get the $\omega=-\frac{3}{2}$ Brans-Dicke theory.
Thus, $T^{\mu}_{\,\,\,\nu}$ in (\ref{kdr}) or (\ref{kel}), together with $\Box{\phi}=0$,
forms an example of a generalized vacuum Brans-Dicke theory (although it can also be equipped by
other equations of motion for $\phi$). 
We close with a remark in relation to equation (\ref{utc}). While the energy-momentum tensor
(\ref{utc}) contains an arbitrary function, in the same context, assuming an extra varying cosmological
constant, a unique theory arises \cite{Kofinas:2015sna} whose energy-momentum tensor consists of
the $\omega=-\frac{3}{2}$ Brans-Dicke $T^{\mu}_{\,\,\,\nu}$ with the addition
of the term $-\frac{\phi^{2}}{8\pi}\delta^{\mu}_{\,\,\,\nu}$ (in \cite{Kofinas:2015sna},
$T^{\mu}_{\,\,\,\nu}$ was expressed in terms of $\psi=\ln(\phi/\bar{\Lambda})$).
This is due to that an extra term of the form $F(\phi)\phi_{;\nu}$ appears in (\ref{ask}). Therefore,
a varying cosmological constant extracts a single vacuum theory instead of an infinity of theories here,
and moreover, this theory is an improvement of the $\lambda\rightarrow \infty$ Brans-Dicke theory.
The fact that both $T^{\mu}_{\,\,\,\nu}$, the $\omega=-\frac{3}{2}$ Brans-Dicke and the one in
\cite{Kofinas:2015sna},
are covariantly conserved is possible since these conservations occur on-shell and the Einstein
equations are different in the two cases.

\section{Conclusions} \label{Conclusions}

We have studied the generalization of Brans-Dicke gravity theory, assuming the same
equation of motion $\Box\phi=4\pi\lambda\mathcal{T}$ for the scalar field $\phi$ and a general
energy-momentum tensor for $\phi$ with two derivatives in each term, but we have relaxed
the exact energy conservation of the matter stress tensor $\mathcal{T}^{\mu}_{\,\,\,\,\nu}$.
Exhaustively investigating the restrictions implied by the Bianchi identities, we have
found only three possible interaction terms on the right hand side of the conservation
equation of the form $\mathcal{T}^{\mu}_{\,\,\,\,\nu}\phi_{;\mu}$,
$\mathcal{T}\phi_{;\nu}$, $\mathcal{T}_{;\nu}$. One or two arbitrary functions of $\phi$
parametrize the consistent theories, which therefore lack naturalness and theoretical
significance. There are only three unique theories, each containing one interaction term,
which are unambiguously determined and form the predominant completions of Brans-Dicke theory.
These theories contain new dimensionfull or dimensionless parameters which appear as
integration constants through the integration procedure. When these parameters take
suitable values, the theories reduce to the standard Brans-Dicke.

In the absence of matter other than $\phi$, the family of vacuum theories obeying the free
wave equation for the scalar field is parametrized by an arbitrary function of $\phi$.
The three vacuum theories arising from the zero matter limit of the complete matter theories
possess prominent position in this family. The subclass of vacuum theories with identically
covariantly conserved energy-momentum tensors is also found, it is parametrized by one
free function of $\phi$, and these theories can be supplemented by any equation of motion for
the scalar field.

It would be interesting to study the cosmology and the black hole solutions of the three complete
Brans-Dicke theories, to investigate the stability issues, or to confront with the solar system
bounds and restrict the parameters encountered.

\begin{acknowledgments}
I wish to thank Alex Kehagias for useful discussions.
\end{acknowledgments}


\end{document}